\documentclass[twocolumn]{aastex63}

\graphicspath{{./}{figures/}}
\usepackage{amssymb}
\usepackage{mathtools} 

\shorttitle{Swift J1818.0$-$1607 spectropolarimetric properties}
\shortauthors{Lower et al.}

\begin{document}

\title{Spectropolarimetric properties of Swift J1818.0$-$1607: a 1.4\,s radio magnetar}

\correspondingauthor{Marcus E. Lower}
\email{mlower@swin.edu.au}

\author[0000-0001-9208-0009]{Marcus E. Lower}
\affiliation{Centre for Astrophysics and Supercomputing, Swinburne University of Technology, PO Box 218, Hawthorn, VIC 3122, Australia}
\affiliation{CSIRO Astronomy and Space Science, Australia Telescope National Facility, Epping, NSW 1710, Australia}

\author[0000-0002-7285-6348]{Ryan M. Shannon}
\affiliation{Centre for Astrophysics and Supercomputing, Swinburne University of Technology, PO Box 218, Hawthorn, VIC 3122, Australia}
\affiliation{OzGrav: The ARC Centre of Excellence for Gravitational-wave Discovery, Hawthorn VIC 3122, Australia}

\author[0000-0002-7122-4963]{Simon Johnston}
\affiliation{CSIRO Astronomy and Space Science, Australia Telescope National Facility, Epping, NSW 1710, Australia}

\author[0000-0003-3294-3081]{Matthew Bailes}
\affiliation{Centre for Astrophysics and Supercomputing, Swinburne University of Technology, PO Box 218, Hawthorn, VIC 3122, Australia}
\affiliation{OzGrav: The ARC Centre of Excellence for Gravitational-wave Discovery, Hawthorn VIC 3122, Australia}

\begin{abstract}
The soft-gamma repeater Swift J1818.0$-$1607 is only the fifth magnetar found to exhibit pulsed radio emission. Using the Ultra-Wideband Low receiver system of the Parkes radio telescope, we conducted a 3\,h observation of Swift J1818.0$-$1607. 
Folding the data at a rotation period of $P = 1.363$\,s, we obtained wideband polarization profiles and flux density measurements covering radio frequencies between 704-4032\,MHz. 
After measuring, and then correcting for the pulsar's rotation measure of $1442.0 \pm 0.2$\,rad\,m$^{-2}$, we find the radio profile is between 80-100 per cent linearly polarized across the wide observing band, with a small amount of depolarization at low frequencies that we ascribe to scatter broadening.
We also measure a steep spectral index of $\alpha = -2.26^{+0.02}_{-0.03}$ across our large frequency range, a significant deviation from the flat or inverted spectra often associated with radio-loud magnetars.
The steep spectrum and temporal rise in flux density bears some resemblance to the behaviour of the magnetar-like, rotation-powered pulsar PSR J1119$-$6127. This leads us to speculate that Swift J1818.0$-$1607 may represent an additional link between rotation-powered pulsars and magnetars. 
\end{abstract}

\keywords{Magnetars (992) -- Neutron stars (1108) -- Pulsars (1306) -- Radio pulsars (1353)}

\section{Introduction} \label{sec:intro}

Magnetars are a rare class of relatively slow rotating neutron star that are inferred to possess some of the strongest magnetic fields in the Universe.
Until recently, only 4 of the 23 confirmed magnetars \footnote{\url{http://www.physics.mcgill.ca/~pulsar/magnetar/main.html}}~\citep{Olausen2014} were seen to exhibit pulsed radio emission~\citep{Camilo2006, Camilo2007a, Levin2010, Eatough2013, Shannon2013}. 
Unlike standard rotation-powered pulsars, the radio pulses seen from these magnetars have generally flat spectra and display highly variable flux densities over timescales ranging between seconds to months~\citep{Camilo2007c, Lazaridis2008}. 
Their single pulses are often comprised of many burst-like sub-pulses that display a remarkable range of temporal phenomenology. These sub-pulses have drawn comparisons to similar `spiky' emission seen in high magnetic field strength pulsars~\citep{Weltevrede2011}, rotating radio transients~\citep[RRATs;][]{McLaughlin2006}, and fast radio bursts~\citep[FRBs;][]{Pearlman2018}.
Observations covering wide radio frequency bands may shed light on their magnetospheric conditions following outbursts, in particular whether the same processes that produce coherent, highly polarized emission in rotation-powered pulsars is also responsible for pulsed radio emission from magnetars.

\begin{figure*}
    \centering
    \includegraphics[width=\linewidth]{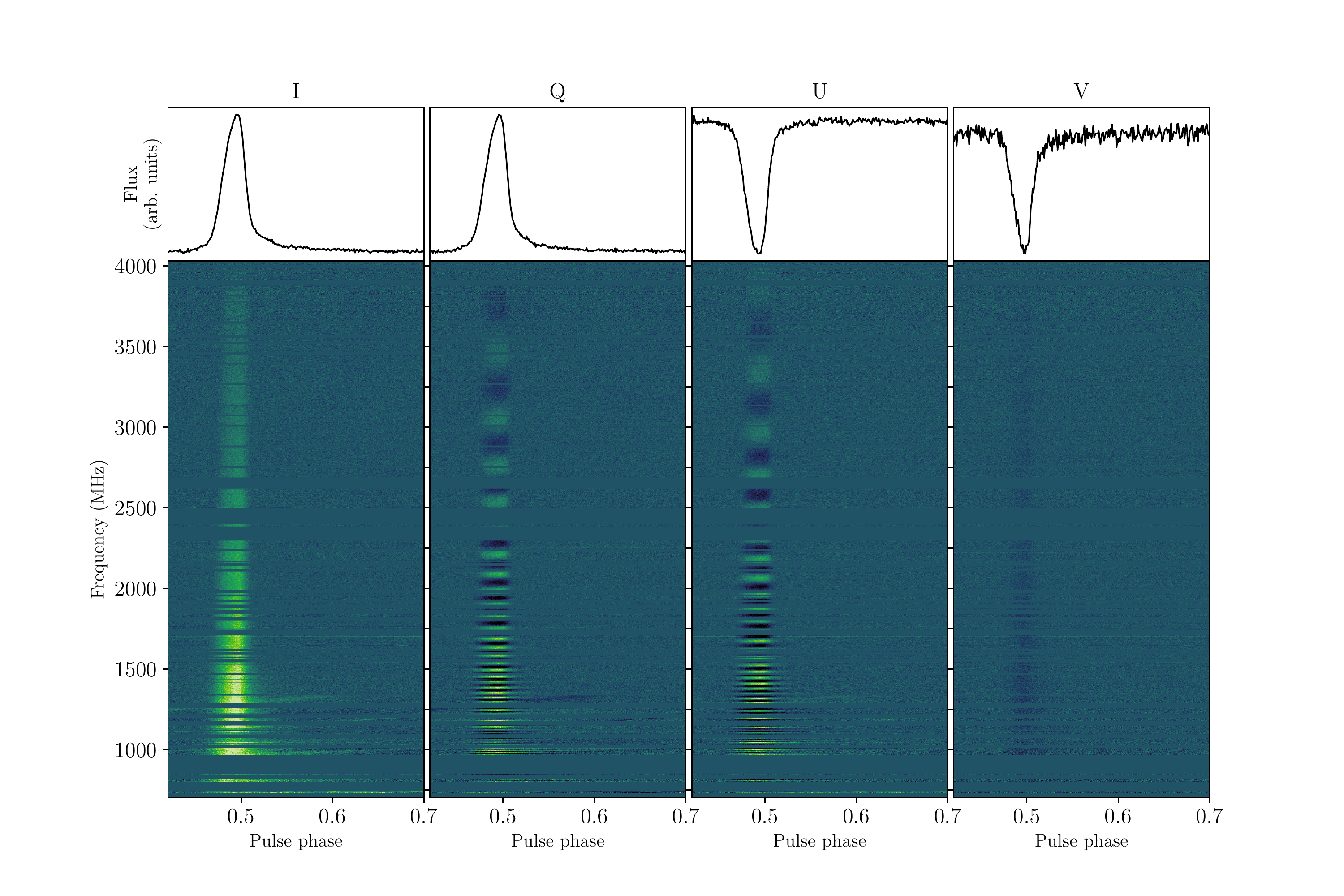}
    \caption{Faraday-corrected average polarization profiles (\textit{top}) and uncorrected, time-averaged polarization spectra (\textit{bottom}) of Swift J1818.0$-$1607. All four Stokes parameters are plotted with 2\,MHz spectral resolution and 0.67\,ms temporal resolution. The large rotation measure of $1442.0 \pm 0.2$\,rad\,m$^{-2}$ is clearly visible in Stokes Q and U. Horizontal gaps in each panel represent frequency channels that were excised due to RFI contamination. Some broadband sweeps of RFI remain visible below 1300\,MHz.}
    \label{fig:spectrum}
\end{figure*}

Recently a fifth radio-bright magnetar was identified.
Swift J1818.0$-$1607 was discovered by the \textit{Swift} space observatory following the detection of a gamma-ray outburst by the Burst Alert Telescope on MJD 58920 (2020-03-12-21:16:47 UT). The burst was quickly localized to an X-ray point source by the on-board X-ray telescope~\citep{Evans2020GCN}. 
Observations by the \textit{Neutron star Interior Composition Explorer} found the source exhibited pulsed X-ray emission with a periodicity of $1.36$\,s~\citep{Enoto2020ATel}. Two days after the initial outburst, highly linearly polarized radio pulsations were detected with a dispersion measure (DM) of $706 \pm 4$\,pc\,cm$^{-3}$ during follow-up observations by the 100-m Effelsberg radio telescope observing in a band centered on 1.37\,GHz~\citep{Karuppusamy2020ATel}.
Continued timing provided an initial measurement of the spin-period derivative, $\dot{P} = 9 \pm 1 \times 10^{-11}$~\citep{Esposito2020}, firmly cementing its status as the fastest rotating, and possibly the youngest magnetar found to date. 
Observations performed at multiple radio wavelengths indicated the magnetar's radio emission has a steep spectral index~\citep{Gajjar2020ATel, Lower2020ATelb}. This is similar to the observed radio spectra of many ordinary, rotation-powered radio pulsars, but significantly differs from the flat or inverted spectra of the four other radio loud magnetars. 
The apparently low surface temperature~\citep{Esposito2020} and lack of coincident supernova remnant, indicate Swift J1818.0$-$1607 may be significantly older than implied by its characteristic age of $240-310$\,yrs, and may represent a transitional link between magnetars and the population of high B-field, rotation-powered pulsars.
In particular, the reported spectral flattening by~\citet{Majid2020ATel} may indicate a possible link to the 2016 magnetar-like outburst of PSR J1119$-$6127~\citep{Majid2017}.

In this letter we report on observations of Swift J1818.0$-$1607 using the Ultra-wideband Low (UWL) receiver system~\citep{Hobbs2020} of the CSIRO 64-m Parkes radio telescope.
Using Bayesian inference techniques, we measured the broadband properties of the time averaged polarization spectrum and analyzed the sample of bright single pulses observed throughout the approximately $3$\,hour-long observation. 
We then compare these results to previous observations of the four other radio loud magnetars and the general pulsar population. Finally, we discuss the potential evolutionary pathways of Swift J1818.0$-$1607.

\section{Observation and analysis} \label{sec:obs}

\begin{figure*}
    \centering
    \includegraphics[width=\linewidth]{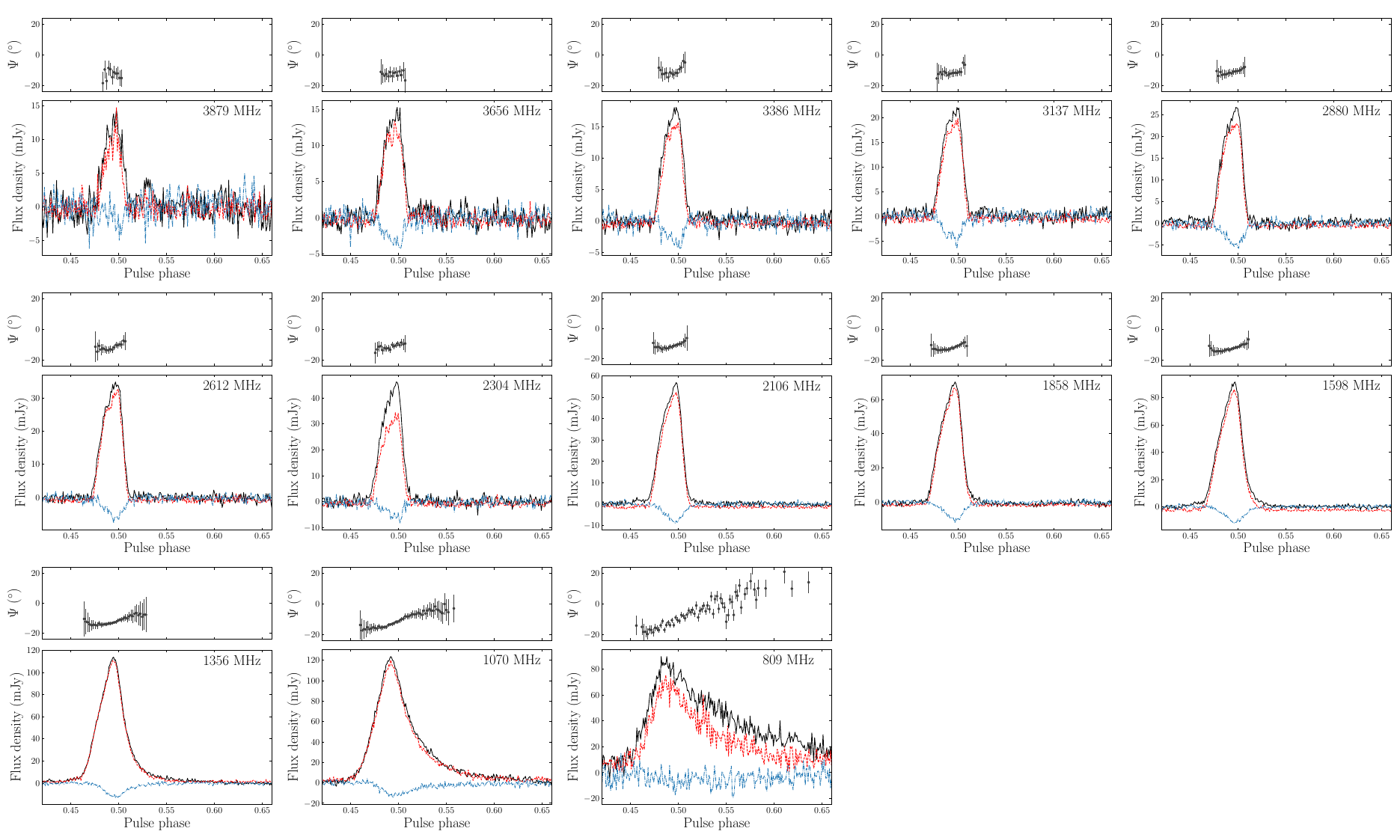}
    \caption{Polarization profiles of Swift J1818.0$-$1607 averaged over 13 sub-bands from 3656\,MHz to 809\,MHz, each covering 256\,MHz of bandwidth. Black represents total intensity, red linear polarization and blue is circular polarization. The linear polarization position angles ($\Psi$) are corrected for the RM $= 1442.0 \pm 0.2$\,rad\,m$^{-2}$ at a reference frequency of 2368\,MHz.}
    \label{fig:pstg_stmp}
\end{figure*}

We conducted a 10473\,s observation of Swift J1818.0$-$1607 on MJD 58939 using the Parkes UWL receiver~\citep{Hobbs2020} under the target of opportunity request PX057 (PI: Lower).
Pulsar search-mode data with 128\,$\mu$s sampling covering the full UWL band from 704-4032\,MHz with full Stokes information were recorded using the {\sc medusa} backend and coherently dedispersed on a channel by channel basis at a DM of $700$\, pc\,cm$^{-3}$ to minimize dispersive smearing of the pulse profile. Note the profiles shown in Figures~\ref{fig:spectrum} and~\ref{fig:pstg_stmp} have been dedispersed using the inferred DM of $706.0$\,pc\,cm$^{-3}$ from Section~\ref{sec:single_pulses}. 
The data were then folded at the pulse period of the magnetar using \texttt{DSPSR}~\citep{vanStraten2011} and 
saved to a \texttt{psrfits}~\citep{Hotan2004} format archive with 1024 phase bins, and 3328 frequency channels with 1\,MHz frequency resolution. 
Approximately 35\,per cent of the 3328 frequency channels were heavily contaminated by radio frequency interference (RFI), and were subsequently excised using the standard \texttt{paz} and \texttt{pazi} tools in \texttt{PSRCHIVE}~\citep{Hotan2004, vanStraten2012}. 
The data were flux and polarization calibrated in the same manner as~\citet{Dai2019}, with the exception that we used the radio galaxy PKS B0407$-$658 as a flux density reference instead of 3C\,218.
Unlike 3C\,218, PKS B0407$-$658 is not resolved by Parkes above $\sim 3$\,GHz, making it a more reliable calibrator for the UWL.
We used an observation of a linearly polarized noise diode prior to observing the magnetar, in addition to on- and off-source observations of PKS B0407$-$658 taken on MJD 58638 to measure the noise diode brightness and to correct the phase and absolute gain of the system.
We note that any leakage terms were not corrected for, which may be of order 5 per cent toward the top of the band.

\subsection{Profile phenomenology and flux density}\label{sec:phenom}


\begin{table}
\centering
\caption{Scatter broadening ($\tau_{\mathrm{sc}}$), period-averaged flux density ($S_{\nu}$) measurements, and fractional linear ($\langle L/I \rangle$) and circular ($\langle |V|/I \rangle$) polarization of each 256\,MHz sub-band. The uncertainties denote the 68\% confidence intervals. Only upper limits are set on the scattering timescale at frequencies above 2880\,MHz and are with 68\% confidence.}
\label{tbl:fits_n_flux}
\begin{tabular}{lcccc}
\hline
\hline
Frequency & $\tau_{\mathrm{sc}}$ & $S_{\nu}$ & $\langle L/I \rangle$ & $\langle |V|/I \rangle$ \\
(MHz) & (ms) & (mJy) & & \\
\hline
3879 & $\lesssim 3$    & $0.31 \pm 0.03$ & 0.73 & 0.19 \\
3656 & $\lesssim 3$    & $0.33 \pm 0.01$ & 0.94 & 0.18 \\
3386 & $\lesssim 3$    & $0.41 \pm 0.01$ & 0.86 & 0.18 \\
3137 & $\lesssim 1.9$  & $0.52 \pm 0.01$ & 0.88 & 0.20 \\
2880 & $0.8 \pm 0.5$   & $0.62 \pm 0.01$ & 0.88 & 0.18 \\ 
2612 & $1.2 \pm 0.6$   & $0.82 \pm 0.01$ & 0.92 & 0.16 \\
2304 & $2.8 \pm 0.5$   & $1.11 \pm 0.02$ & 0.73 & 0.12 \\
2106 & $3.5 \pm 0.2$   & $1.40 \pm 0.01$ & 0.92 & 0.11 \\
1858 & $5.3 \pm 0.2$   & $1.79 \pm 0.02$ & 0.97 & 0.12 \\
1598 & $8.8 \pm 0.2$   & $2.53 \pm 0.01$ & 0.93 & 0.11 \\
1356 & $16.8 \pm 0.2$  & $3.72 \pm 0.1$  & 0.93 & 0.10 \\
1070 & $38.6 \pm 0.5$  & $6.0 \pm 0.1$   & 0.83 & 0.16 \\
809  & $186^{+7}_{-6}$ & $11.8 \pm 0.6$  & 0.52 & 0.18 \\
\hline
\end{tabular}
\end{table}

Dynamic spectra showing the four Stokes parameters across the continuous 704-4032\,MHz UWL band are displayed in Figure~\ref{fig:spectrum}. The pulse profile shows clear evidence of a steep negative gradient in flux density, and can be described as the superposition of two Gaussian components ($G_{1}$ and $G_{2}$ hereafter). The narrower $G_{2}$ component appears brighter toward the lower end of the UWL band, indicating it has a steeper spectral index than $G_{1}$. 


\begin{figure}
    \centering
    \includegraphics[width=\linewidth]{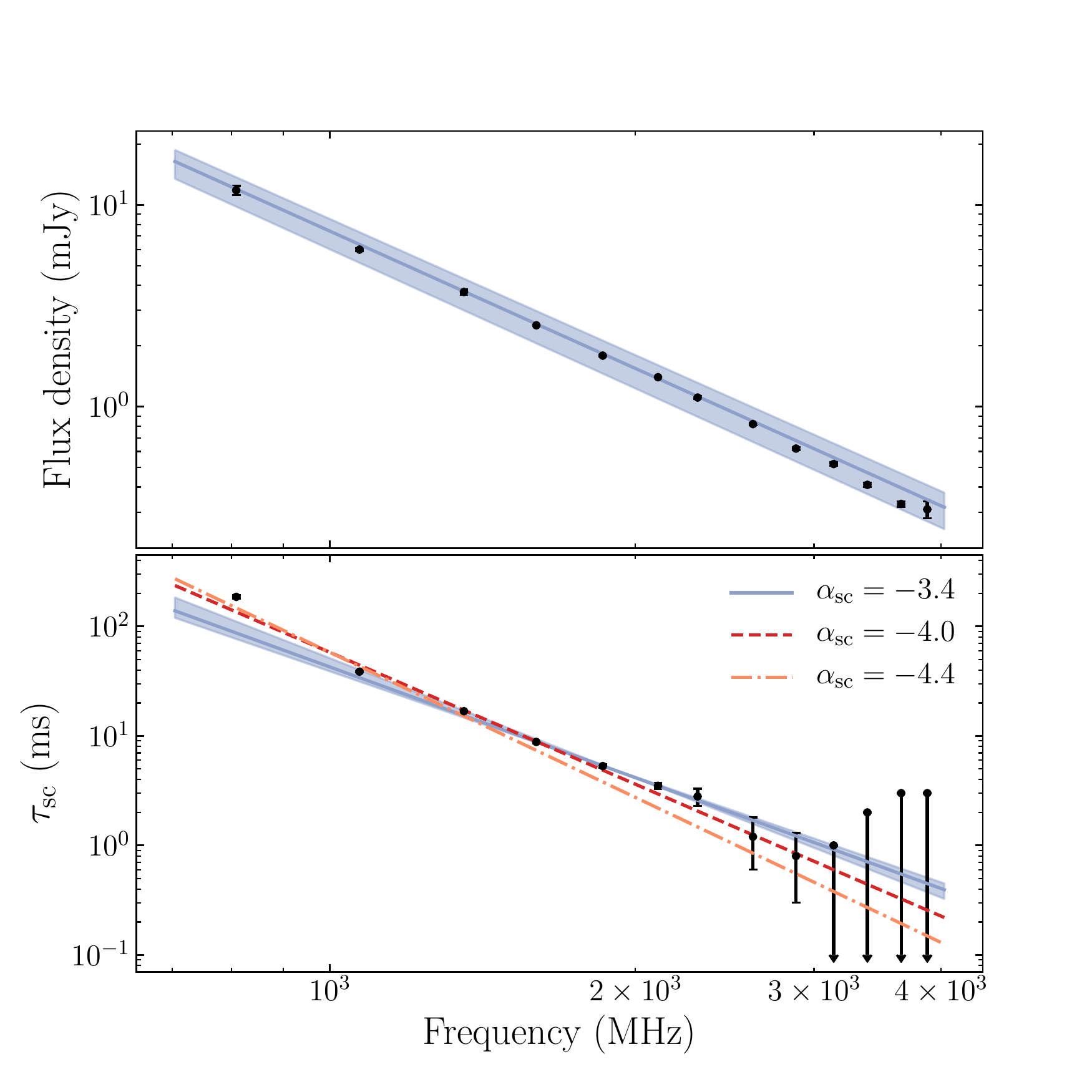}
    \caption{Period-averaged flux density (\textit{top}) and scattering timescale (\textit{bottom}) as a functions of frequency. The blue solid lines indicates the median fit while the shaded region is bounded by the 68\% confidence intervals. Dashed red and dash-dotted orange lines correspond to scattering indices of $-4$ and $-4.4$ respectively.}
    \label{fig:fits}
\end{figure}

We further analyzed  the profile by dividing the data into 13 sub-bands, each having 256\,MHz of bandwidth. These sub-bands were then averaged in frequency and polarization before being fit with a two-component Gaussian profile convolved with a one-sided exponential pulse broadening function
\begin{equation}
    f(t) = \sum_{i=1}^{2} \frac{1}{\sqrt{2\pi\sigma_{i}^{2}}} e^{-(t - \mu_{i})^{2}/2\sigma_{i}^{2}} \circledast e^{-t/\tau_{\mathrm{sc}}},
\end{equation}
where $\mu_{i}$ and $\sigma_{i}$ are the centroids and widths of the $i$-th Gaussian component, $\circledast$ indicates a convolution and $\tau_{\mathrm{sc}}$ is the scattering timescale.
The resulting posterior probability distributions were sampled using the \texttt{bilby} software package~\citep{Ashton2019} as a front-end to the \texttt{dynesty} nested sampling algorithm~\citep{Speagle2019}. 
Initially we fit the sub-bands assuming uniform priors on the widths of the profile components $G_{1}$ and $G_{2}$. 
However, we found the component widths were highly covariant with the scattering timescale, to the point where we could only recover upper-limits for scattering in sub-bands above 2106\,MHz.
As the profile width does not appear to undergo significant evolution with frequency, aside from scatter broadening, we re-fit the sub-banded data assuming Gaussian priors of $\pi(\sigma_{1}) \sim \mathcal{N}(8\,\mathrm{ms}, 1\,\mathrm{ms})$ and $\pi(\sigma_{2}) \sim \mathcal{N}(7\,\mathrm{ms}, 1\,\mathrm{ms})$ for the widths of $G_{1}$ and $G_{2}$ respectively.

The resulting scattering timescale and period-averaged flux density -- measured by averaging the best-fit template for each sub-band in pulse phase -- are presented in Table~\ref{tbl:fits_n_flux}.
We measure a scattering timescale referenced to 1\,GHz of $\tau_{\mathrm{sc,1\,GHz}} = 42^{+9}_{-3}$\,ms, with a scattering index of $\alpha_{\mathrm{sc}} = -3.4^{+0.3}_{-0.2}$. 
Similar but less well constrained values of $\alpha_{\mathrm{sc}} = 3.6^{+0.8}_{-1.1}$ and $\tau_{\mathrm{sc,1\,GHz}} = 41^{+19}_{-18}$\,ms were obtained when we used uniform priors on the widths of $G_{1}$ and $G_{2}$.
In either case, the scattering timescale is consistent with the expected value of $62\pm 30$\,ms from the NE2001 galactic electron density model at 1\,GHz~\citep{Cordes2002}.
While the scattering index is smaller than the expected value of $\alpha_{\mathrm{sc}} = -4$ or $\alpha_{\mathrm{sc}} = -4.4$ expected from Kolmogorov turbulence, they are consistent with the scattering indices of many other pulsars~\citep[see for example][]{Geyer2017}.
We also fit the period-averaged flux density spectrum using a simple power-law function, $S_{\nu} \propto \nu^{\alpha}$, obtaining a spectral index of $\alpha = -2.26^{+0.02}_{-0.03}$. 
The fits to the spectral index and scattering timescale are plotted in Figure~\ref{fig:fits}.
The reduced $\chi^{2}$ for the scattering relation shown in Figure~\ref{fig:fits} is $13.8$.
We attribute the high value to overestimation of the scattering timescale in the RFI-affected 809\,MHz band.
Removing the 809\,MHz data point confirms this suspicion, as refitting the scattering relation returns a consistent scattering index of $\alpha = -3.6^{+0.4}_{-0.3}$ and a reduced $\chi^{2}$ of $0.6$.

\subsection{Polarimetry}
 Figure~\ref{fig:spectrum} clearly shows the linear polarization has undergone significant Faraday rotation, as evidenced by the large number of changes in sign for Stokes Q and U. Following the Bayesian methodology presented in~\citet{Bannister2019}, we measured the phase averaged rotation measure (RM) of the magnetar by directly fitting Stokes Q and U as a function of frequency, obtaining a value of $1442.0 \pm 0.2$\,rad\,m$^{-2}$ ($68$\,per cent confidence interval). Note, this measurement does not include corrections for the ionosphere which can often exceed our measurement uncertainty. At Parkes, the ionospheric contribution is typically between $-0.2$ to $-2.0$\,rad\,m$^{-2}$~\citep{Han2018}.

To better visualize the polarization profiles, we plot the averaged polarization pulse profiles at 13 frequencies in Figure~\ref{fig:pstg_stmp}, along with the linear polarization position angle for each sub-band. 
The pulse profile is almost than 90\,per cent linearly polarized across most of the UWL band, although a small amount of circular polarization is also present.
Apparent depolarization due to scatter broadening~\citep{Li2003} is evident below 1356\,MHz.
Slight variations in the fractional linear and circular polarizations listed in Table~\ref{tbl:fits_n_flux} likely result from a combination of noise and polarization impurities in the receiver system.
The apparent depolarization in the 2304\,MHz band is an artefact of residual RFI from wireless communications contaminating the narrow strip of non-excised channels between 2380 and 2400\,MHz.
Additionally, the lack of polarisation in the `bump' visible in the off-pulse noise of the 3879\,MHz sub-band suggests this feature is likely to be residual impulsive RFI, not an additional profile component.
There is a slight upward slope in the linear polarization position angle (PA), with little frequency dependent evolution except for scatter-induced smearing at lower frequencies.

\subsection{Single pulses} \label{sec:single_pulses}

To analyse the single pulses from the magnetar, we created single pulse archives from the original \texttt{psrfits} search-mode filterbank.
We then performed a boxcar search for single pulses on copies of these archives where all frequency channels outside the 1300-2500\,MHz band had been excised to minimize confusion with RFI. 
We limited this search to only the on-pulse region of each archive. 
Applying a maximum boxcar width of 85\,ms and threshold S/N of 7, we find 5052 of the 7008 single pulse archives contained a single pulse candidate that met our criterion with a median S/N of 13.8.
Upon visual inspection, we found the single pulses typically consist of 1-3 `spiky' sub-pulses with similar phenomenology to single pulses seen from the four other radio loud magnetars. 
We did not observe any single pulses emitted at rotational phases outside the `on-pulse' region represented by the integrated profiles in Figure~\ref{fig:pstg_stmp}, nor evidence of sporadic pulses from the additional profile component reported by~\citet{Maan2020ATel}.
Occasional gaps or nulls in emission were seen throughout the observation. 
Similar behaviour has been reported in observations of the galactic centre magnetar SGR J1745$-$2900~\citep{Yan2018}. 
However, it is not clear whether the gaps we observed represent true nulls, where the radio emission mechanism completely shuts off, or if the radio pulses during these rotations were simply below the detection threshold of the receiver. 

\begin{figure}
    \centering
    \includegraphics[width=\linewidth]{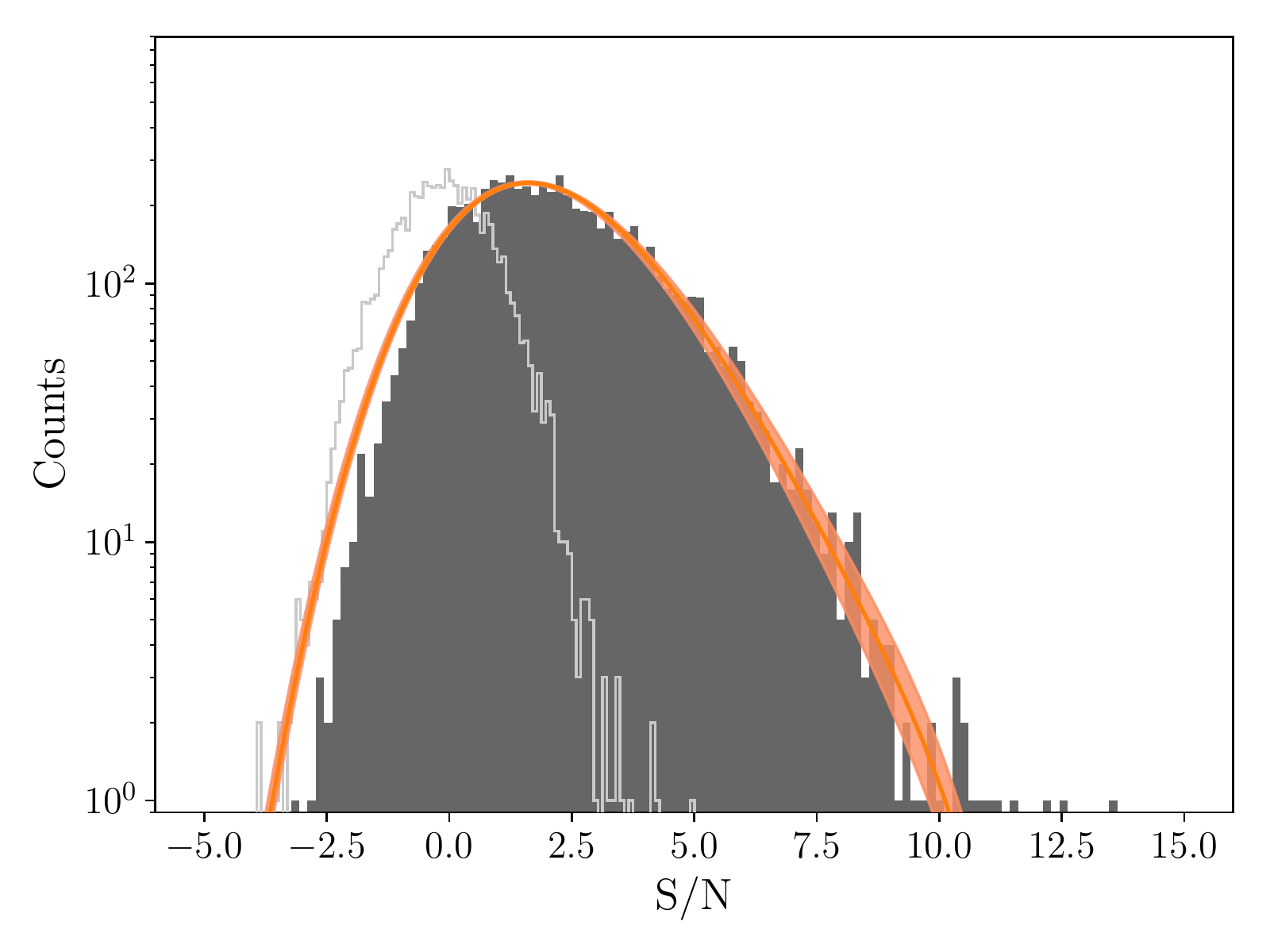}
    \caption{Matched-filter S/N distribution for the frequency-averaged single pulses (dark grey), scaled such that the off-pulse noise (light grey) has zero mean and unit variance. The orange line and shading are the median log-normal convolved with a Gaussian fit to the data and associated $68\%$ confidence intervals.}
    \label{fig:hist}
\end{figure}

We measured the flux density of the on- and off-pulse regions of each single-pulse archive using the \texttt{psrflux} tool from \texttt{PSRCHIVE} by cross-correlating the data with a scatter-broadened Gaussian template. 
Both the on- and off-pulse flux density measurements were then converted to units of matched-filter S/N by scaling each measurement by a factor of 1.4 -- the scale factor needed to scale the off-pulse distribution such that it has a mean of zero and variance of one. 
The resulting on- and off-pulse S/N distributions are shown in Figure~\ref{fig:hist}. 
We note this definition of S/N is different to the one used in the earlier single pulse search, which was a top-hat S/N used to place quantitative constraints on the number of single pulses we detected.
Negative S/N ratios can be attributed to the on-pulse flux being below zero due to fluctuations in the baseline.
The on-pulse distribution is well described by a log-normal with a log-mean of $1.925 \pm 0.003$ and width of $0.25 \pm 0.01$ that has been convolved with a Gaussian distribution with zero mean and unit variance.
This distribution width is typical of the rotation-powered pulsar population as a whole~\citep{Burke-Spolaor2012}.
While there are some outliers, the lack of a power-law tail in the distribution indicates no giant pulses were detected during our observation, contradictory to the claim by~\citet{Esposito2020} that the single pulses are dominated by sporadic giant pulses. 
It is possible their giant pulse detections originated from the transient profile component seen in early observations by~\citet{Maan2020ATel}, which had disappeared sometime prior to our observation with Parkes.

\begin{figure}
    \centering
    \includegraphics[width=\linewidth]{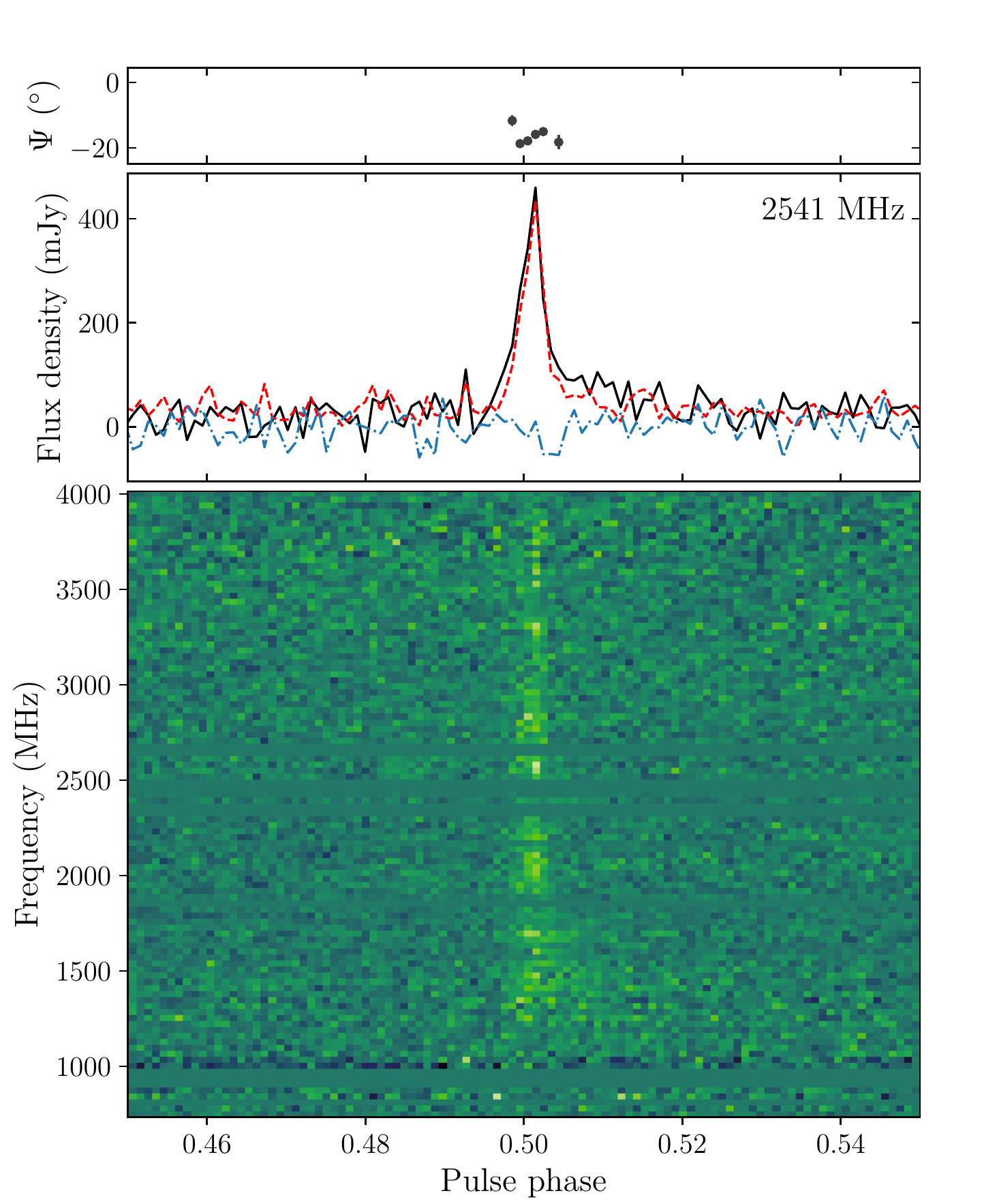}
    \caption{A single pulse from Swift J1818.0$-$1607. The top and middle panels show the position angle and integrated polarization profile. The bottom panel shows the waterfall diagram of the pulse dedispersed at a DM = $707.3 \pm 0.2$\,pc\,cm$^{-3}$ with 0.67\,ms time resolution and 16\,MHz spectral resolution.}
    \label{fig:pulse}
\end{figure}

The narrow widths of magnetar single pulses and sub-pulses enable high-accuracy DM measurements, particularly when observed across large bandwidths.
For example, the bright single pulse shown in Figure~\ref{fig:pulse} returned a structure-optimized DM of $707.3 \pm 0.2$\,pc\,cm$^{-3}$. 
Repeating this for the brightest 215 single pulses in our sample, we find the distribution of structure-optimised DMs is well described by a Gaussian with a mean of $706.0$\,pc\,cm$^{-3}$ and a standard deviation of $2.6$\,pc\,cm$^{-3}$.
From this, we estimated the magnetar's DM to be $706.0 \pm 0.2$\,pc\,cm$^{-3}$ where the uncertainty is derived from the standard deviation of the DM distribution $\sigma_{\mathrm{DM}} = 2.6/(215 -2)^{1/2}$\,pc\,cm$^{-3}$.
The variations in DM are more likely to have resulted from systematic errors in the structure-optimization algorithm combined with the variable number of sub-pulses in each pulse as opposed to short-timescale variations in the local environment of the magnetar. 
Long-term monitoring over year-long timescales will reveal if Swift J1818.0$-$1607 experiences DM variations similar to those seen in repeating FRBs~\citep[e.g.][]{Hessels2019}.

Using the NE2001~\citep{Cordes2002} and YMW16~\citep{Yao2017} galactic free electron density models, the distance to the magnetar is estimated to be either $8.1 \pm 1.6$\,kpc (NE2001) or $4.8$\,kpc (YMW16), where the uncertainty is dominated by the model chosen.

From our measurements of both the RM and DM, we can estimate the average parallel magnetic field strength along the line of sight to the magnetar using the equation $B_{\parallel} = 1.2 \mathrm{RM}/\mathrm{DM}$, where $B_{\parallel}$ is in units of $\mu$G, and the RM and DM are in their usual units (rad\,m$^{-2}$ and pc\,cm$^{-3}$). Our measured value of 2.5\,$\mu$G is fairly typical of line-of-sight $B_{\parallel}$ measurements from pulsars within the galactic plane~\citep{Han2018}

\section{Discussion} \label{sec:discussion}

In general, the pulsed radio emission from Swift J1818.0$-$1607 shares a lot of the same phenomenology seen in other radio loud magnetars: a high degree of linear polarization, burst-like sub-pulses and extremely variable pulse-to-pulse flux densities.
However, the steep spectral index we measure is more consistent with the spectral indices of many rotation powered pulsars when compared to the flat spectral indices of the four other radio magnetars which typically range between $-0.5$ to $+0.3$~\citep{Lazaridis2008, Torne2015, Dai2019}, making this new magnetar a significant outlier.
Given the DM and location of the magnetar, the effects of diffractive interstellar scintillation are negligible at the UWL observing band. 
For instance, the NE2001 model predicts a scintillation bandwidth of only $3^{+3}_{-1}$\,Hz at 1\,GHz.
Hence the steep spectrum is intrinsic to Swift J1818.0$-$1607.
This indicates that it was premature to assume that all radio magnetars have flat spectra. At the large DMs typical of magnetars, those that have steep radio spectra might be so scatter-broadened as to induce a significant selection effect towards those with flatter spectra.
When compared to the 276 pulsars in~\citet{Jankowski2018} that have spectra best fit by a simple power-law, only $\sim 11$\,per cent of pulsars have steeper spectra than Swift J1818.0$-$1607, while the four other radio magnetars all have spectral indices that are flatter than $\sim 94$\,per cent of their sample.
Hence, Swift J1818.0$-$1607 may be an example of the diversity that could exist in the wider, as-of-yet undetected radio magnetar population. 
The spectral properties could also be related to the magnetar possessing a less evolved magnetic field structure due to its youth.

Assuming Swift J1818.0$-$1607 was born rapidly rotating ($P \sim 10$\,ms) and its spin-down is dominated by magnetic dipole radiation (braking index = 3), measurements of its spin and spin-down place its characteristic age between only $240$-$310$\,yrs~\citep{Champion2020ATel, Hu2020ATel, Esposito2020}, the second smallest of any pulsar after SGR J1806$-$20~\citep{Mereghetti2005}. 
However, given large amount of uncertainty surrounding neutron star rotation periods at birth and the diversity in measured pulsar (and magnetar) braking indices, its true age is likely to be significantly different than the inferred spin-down age.
Indeed the period derivatives of magnetars can change by large factors within just a few years~\citep[see for example][]{Scholz2017}.
A more accurate kinematic age could be inferred from associating the magnetar to a progenitor supernova remnant, combined with a proper-motion measurement from very-long baseline interferometry.
However, we find there are no catalogued supernova remnants or pulsar-wind nebula co-located with its position~\citep{Green2019}. 
The two closest supernova remnants (G014.3$+$0.1 and G014.1$-$0.1) are approximately $19$\,arcmin and $27$\,arcmin away from the position of the magnetar on sky (Galactic coordinates: $l=14.8$\,$^{\circ}$, $b=-0.14$\,$^{\circ}$) respectively, making an association highly unlikely.
The lack of an associated supernova remnant is not too surprising, as only eight of the twenty-three known magnetars have claimed associations. 
Additionally, the strong spin-down powered wind from new-born magnetars can accelerate the remnant expansion to the point that only anomalously diffuse shells, or no remnant at all, remains on century-long timescales~\citep{Duncan1992}.
If the progenitor supernova remnant has not been dissipated, then deep radio and X-ray imaging may be able to detect it.

Alternatively, we speculate the steep spectrum and its unusually faint X-ray luminosity of $7 \times 10^{34}$\,ergs\,s$^{-1}$~\citep{Esposito2020}\footnote{As noted in~\citet{Esposito2020}, the quoted X-ray luminosity assumes the smaller, YMW16 DM distance to the magnetar, and that a larger source distance (as implied by the NE2001 model) may yield a more normal luminosity.} may be evidence this new magnetar was initially born as a rotation powered pulsar that obtained the rotational properties of a magnetar over time, similar to what is predicted for PSR J1734$-$3333~\citep{Espinoza2011}. 
Such evolution can occur if the magnetic and spin axes underwent rapid alignment over time~\citep{Johnston2017}, or if the pulsar underwent an extended period of magnetic field growth after the surface magnetic field was initially buried due to fall-back accretion~\citep[e.g.][]{Ho2015}. 

If the properties of Swift J1818.0$-$1607 are the result of rapid magnetic and spin axes alignment,  we would expect the PA to be consistent with that of an aligned rotator.
There is some evidence magnetars tend toward aligned spin and magnetic axes. Both 1E 1547.0$-$5408 and PSR J1622$-$4950 have PA swings that are consistent with being aligned rotators~\citep{Camilo2008, Levin2012}. This is further backed up by the wide radio profiles, and low pulsed X-ray fractions of these two magnetars~\citep{Halpern2008, Camilo2018}.
There is some ambiguity as to whether the spin and magnetic axes of XTE J1810$-$197 are aligned or orthogonal, as~\citet{Camilo2007c} found both scenarios adequately describe the PA swing across its main pulse and inter-pulse.
Conversely, \citet{Kramer2007} found that an offset dipole described by two separate rotating vector models~\citep[RVMs,][]{Radhakrishnan1969} could also describe its PA behaviour, and speculated it may be evidence for XTE J1810$-$197 having a multi-pole magnetic field. Additionally,~\citet{Dai2019} observed distinctly non-RVM PA variations following its 2018 outburst. 
For Swift J1818.0$-$1607, the flat PA in the higher-frequency panels of Figure~\ref{fig:pstg_stmp} is broadly consistent with the RVM for a dipole magnetic field. 
However, the narrow pulse duty cycle makes it difficult to constrain the star's magnetic geometry, as the relatively flat PA could be consistent with either nearly aligned magnetic and spin axes, or a large offset between the magnetic axis and our line-of-sight.
Given the radio profiles of magnetars evolve over the weeks to months following an outburst~\citep{Kramer2007, Dai2019}, it may be possible to measure the magnetic geometry of Swift J1818.0$-$1607 in the future.

Pulsars that experienced fall-back accretion soon after their birth can undergo apparent magnetic field growth as their magnetic fields diffuse to the surface over time~\citep[see for example][]{Muslimov1995}.
This can result in a seemingly `normal' rotation-powered, young pulsar obtain magnetar-like rotational properties within $\sim1$-$10$\,kyr~\citep{Ho2015}.
If Swift J1818.0$-$1607 is a result of this evolutionary path, then we may expect it to show similar radio properties to the high B-field PSRs J1119$-$6127, J1208$-$6238 and J1846$-$0258. 
While PSRs J1846$-$0258~\citep{Gavriil2008} and J1119$-$6127~\citep{Archibald2016} have been observed to undergo magnetar-like outbursts in the past, only PSR J1119$-$6127 has been observed to emit radio pulses. 
Observationally, we can draw parallels between the radio properties of Swift J1818.0$-$1607 and those of PSR J1119$-$6127 during its 2016 outburst. 
Following the initial suppression and re-emergence of radio pulses from PSR J1119$-$6127, multi-band flux measurements found the pulsar possessed a steeper radio spectrum than its nominal $\alpha = -1.4 \pm 0.2$, with values of $\alpha$ ranging between $-2.2 \pm 0.2$ to $-1.9 \pm 0.2$~\citep{Majid2017}. 
Later observations found its radio spectrum had undergone spectral flattening to a more magnetar-like spectral index of $-0.52 \pm 0.06$ over the months following the outburst~\citep{Pearlman2016ATel}. 
The flux density of PSR J1119$-$6127 also underwent a factor of $5$ increase in two weeks after the outburst before recovering back to its normal levels~\citep{Dai2018}. 
In addition to having a comparably steep post-outburst spectral index, Swift J1818.0$-$1607 appears to have also undergone a similar radio brightening, as the flux densities at 1356\,MHz and 1598\,MHz in Table~\ref{tbl:fits_n_flux} are a factor of $5$-$12$ times higher than measurements at similar observing frequencies two weeks prior to our Parkes UWL observation~\citep{Karuppusamy2020ATel, Esposito2020, Lower2020ATelb}.
The refractive modulation timescale is expected to be very long (years) and the modulation index to be low~\citep{Cordes2002}. Thus the increase in flux density cannot be ascribed to refractive effects.
If the current outburst of Swift J1818.0$-$1607 continues to proceed in a similar manner to the 2016 outburst of PSR J1119$-$6127, then we may expect the steep spectral index to undergo a similar flattening and for the flux density to decay to a more steady state over the coming months.
A more recent spectral index measurement of $\alpha = -1.9 \pm 0.2$ from multi-band observations~\citep{Majid2020ATel} suggests some amount of spectral-flattening may have already occurred.
Continued monitoring with multi-band and wide-bandwidth receiver systems will either confirm the spectral index is flattening toward a more magnetar-like value, or is simply fluctuating about some mean value.
Additionally, a measurement of the braking index would allow us to understand the future spin and magnetic field evolution of the magnetar and potentially confirm or rule out a rotation-powered pulsar origin.


\section*{Acknowledgements}

The Parkes radio telescope is part of the Australia Telescope National Facility which is funded by the Australian Government for operation as a National Facility managed by CSIRO.
We acknowledge the Wiradjuri people as the traditional owners of the Observatory site.
This work made use of the OzSTAR national HPC facility, which is funded by Swinburne University of Technology and the National Collaborative Research Infrastructure Strategy (NCRIS).
This work was supported by the Australian Research Council (ARC) Laureate Fellowship FL150100148 and the ARC Centre of Excellence CE170100004 (OzGrav).
MEL receives support from the Australian Government Research Training Program and CSIRO Astronomy and Space Science.
RMS is supported through ARC Future Fellowship FT190100155.
We thank Shi Dai for useful discussions and suggestions on calibrating the UWL data.
We are grateful to the ATNF staff, in particular James Green, for allocating us time to perform this observation.
We also thank the anonymous referee for their helpful comments and suggestions.
This work made use of the Astronomer's Telegram and NASA's Astrophysics Data Service. 

\software{\texttt{bilby} \citep{Ashton2019}, \texttt{clfd} \citep{Morello2019}, \texttt{cmasher} \citep{cmasher}, \texttt{DM\_phase} \citep{DM_phase}, \texttt{dspsr} \citep{vanStraten2011}, \texttt{matplotlib} \citep{matplotlib}, \texttt{numpy} \citep{numpy}, \texttt{psrchive} \citep{Hotan2004, vanStraten2012}.}

\bibliographystyle{aasjournal}
\bibliography{main}

\end{document}